\documentclass{nature}
\usepackage{graphicx}
\usepackage{epstopdf}
\usepackage{dcolumn}
\usepackage{bm}
\usepackage{amssymb}
\usepackage{lineno}
\newcommand{\beq}{\begin{eqnarray}}
\newcommand{\eeq}{\end{eqnarray}}

\title{Photoexcitation Circular Dichroism in Chiral Molecules}
\author{S. Beaulieu$^{1,2}$,
A. Comby$^{1}$,
D. Descamps$^{1}$,
B. Fabre$^{1}$,
G. A. Garcia$^{3}$,
R. G\'eneaux$^{4}$,
A. G. Harvey$^{5}$,
F. L\'egar\'e$^{2}$,
Z. Ma\v{s}\'{i}n$^{5}$,
L. Nahon$^{3}$,
A. F. Ordonez$^{5,6}$,
S. Petit$^{1}$,
B. Pons$^{1}$,
Y. Mairesse$^{1}$,
O. Smirnova$^{5,6}$,
V. Blanchet$^{1}$}

\begin{document}
\maketitle
\begin{affiliations}
\item Universit\'e de Bordeaux - CNRS - CEA, CELIA, UMR5107, F33405 Talence, France
\item Institut National de la Recherche Scientifique, Varennes, Quebec, Canada
\item Synchrotron Soleil, l'orme des Merisiers, BP48, St Aubin, 91192 Gif sur Yvette, France
\item LIDYL, CEA, CNRS, Universit{\'e} Paris-Saclay, CEA Saclay, 91191 Gif-sur-Yvette France. 
\item Max-Born-Institut, Max-Born-Str. 2A, 12489 Berlin, Germany
  \item Technische Universit{\"a}t Berlin,
Ernst-Ruska-Geb{\"a}ude, Hardenbergstr. 36 A, 10623, Berlin, Germany
\end{affiliations}

%\date{\today}
\begin{abstract}
Chirality is ubiquitous in nature and fundamental in science, from particle physics to  metamaterials.
The most established technique of chiral discrimination - photoabsorption circular dichroism  -   relies on the magnetic properties of a chiral medium and yields an extremely weak chiral response.  We propose and demonstrate  a new, orders of magnitude more sensitive type of circular dichroism in neutral molecules: photoexitation circular dichroism. It  does not rely on weak magnetic effects, 
but takes advantage of the coherent helical motion of bound electrons excited by ultrashort circularly polarized light.
 It results in an ultrafast chiral response and the efficient excitation of a macroscopic chiral density in an initially isotropic ensemble of randomly oriented chiral molecules.  We probe this excitation without the aid of further chiral interactions using linearly polarized laser pulses. Our time-resolved study of vibronic chiral dynamics opens a way to the efficient initiation, control and monitoring of chiral chemical change in neutral molecules at the level of electrons.
\end{abstract}

%\pacs{34.80.Gs, 34.80.Ht}
\maketitle
The macro-world gives us many examples of chiral dynamics created by helical structures which convert rotations in a plane into translational motion orthogonal to it, from the Archimedes screw to plane propellers and household fans. %Inverting the structure's chirality reverses the direction of the helical current it creates. 
In the micro-world, the electrons bound inside chiral molecules should develop a similar helical motion when excited by planar rotation of the electric field of circularly polarized light.
Electronic excitation by circularly polarized light has been used to distinguish right-handed from left-handed molecules since 1896 \cite{cotton1896}. The technique, called the photoabsorption circular dichroism (CD) \cite{barron_molecular_2004}, is based on the difference in the absorption of left- and right-circularly polarized light in chiral molecules and remains the go-to tool \cite{berova_circular_2000} for analysing properties of  biological molecules, providing indispensable information on their structure, kinetics and thermodynamics, interaction with the environment and with other molecules. However, it does not rely on the helical nature of bound electron currents \cite{footnote}, but uses the helical pitch of the light wave instead. This pitch, given by the wavelength of the absorbed light, $\lambda \gtrsim 2500 \AA$  ($1\AA=10^{-8}$ cm), is barely noticeable on the molecular scale of $\sim 1 \AA$, leading to very weak signals, three to four orders of magnitude less than the light absorption itself. Formally, the chiral-sensitive part of the light-induced excitation requires the excited electrons to respond to both the electric and the magnetic field of the light wave\cite{footnote2}, see Fig.(1a).

Remarkably, in spite of extraordinary recent advances in developing new methods for chiral discrimination that do not rely on the magnetic properties of the medium \cite{ritchie76,powis00,Bowering01,lux12,lehmann13,pitzer13,herwig_imaging_2013,
patterson2013enantiomer,nahon15,nahon16,Yachmenev2016, comby2016relaxation}, none has relied on detecting the helical motion of bound electrons. Is it possible to excite and probe such motion? 
We demonstrate both theoretically and experimentally that  one can (i) induce chiral stereodynamics of bound electrons without the help of magnetic field effects, a new phenomenon we call PhotoeXcitation Circular Dichroism (PXCD);  
(ii) probe it with linearly polarized light without the help of the further chiral interactions that are usually presumed to be a prerequisite for chiral discrimination.
Coherent excitation substitutes further chiral interactions at the probe step by coupling to a different quantum state of the same chiral molecule. 
 
 \section{Exciting chiral dynamics in bound states}
  A hallmark of helical motion of bound electrons is the appearance of an induced dipole orthogonal to the polarization plane of the exciting circular light. We first show that an ultrashort pulse creates 
 such a dipole in a randomly oriented molecular ensemble.  
Let the electric field of the  pulse, rotating in the x-y plane, coherently excites two states (Fig.1b) of a chiral molecule. 
As shown in the Supplementary Information (SI), the orientation-averaged induced dipole  acquires the desired component along
the light propagation direction $z$:
 \beq
 \label{PXCD}
 {d_z^{PXCD}}\propto\sigma[\vec{d}_{01}\times \vec{d}_{02}]\vec{d}_{12}\sin(\Delta E_{21}t),
 \eeq
Here $\vec{d}_{01}$, $\vec{d}_{02}$ and  $\vec{d}_{12}$ are the dipole transition vectors connecting 
the ground $|0\rangle$ and the two excited states $|1\rangle ,|2\rangle$ (Fig. 1b), $\Delta E_{21}$ is 
the energy spacing between the excited states. For more than two states, Eq.(1) will contain the sum over all pairs of excited states $n,m$, leading to oscillations 
at all relevant
frequencies $\Delta E_{nm}$. As a function of time the  induced dipole vector  maps out a helix (Fig. 1b) and the z-component of the helical  current is
\beq
 \label{PXCD_cur}
 {j_z^{PXCD}}\propto \sigma[\vec{d}_{01}\times \vec{d}_{02}]\vec{d}_{12}\Delta E_{21}\cos(\Delta E_{21}t).
 \eeq
%The coherence created between the states $|1\rangle ,|2\rangle$  does not lead to average dipole in $x$ and $y$ directions. However,  ${d_y^{av}}$, ${d_x^{av}}$ arise due to coherence with the ground state (but do not involve chiral observables, see the  SI). As a function of time the averaged dipole vector $\vec d \equiv (d_x^{av},d_y^{av},d_z^{PXCD})$  maps out a helix (Fig. 1 (e)).
Both $d_z^{PXCD}$ and $j_z^{PXCD}$ are quintessential chiral observables (see e.g. \cite{barron1986true,tang2010optical}).  
Indeed, both are proportional to the light helicity $\sigma=\pm1$ and to the triple product of three vectors $[\vec{d}_{01}\times \vec{d}_{02}]\vec{d}_{12}$. This product presents a fundamental measure of chirality: it changes sign upon  reflection and thus has an opposite sign for left and right enantiomers. For randomly oriented non-chiral molecules $d_z^{PXCD}=j_z^{PXCD}=0$.

Eqs.(\ref{PXCD},\ref{PXCD_cur}) lead to the following conclusions. First, the coherent excitation of electronic states leads to a charge displacement in the light propagation direction. Hence, a macroscopic dipole $d_z^{PXCD}$ and the corresponding chiral density are created in the excited states, with a chiral current oscillating out of phase for the two enantiomers.  Second, 
PXCD requires no magnetic or quadrupole effects. Hence, it is  orders of magnitude stronger than standard photoabsorption CD.
%and does not rely on the chiral properties of circularly polarized light. 
While  photoabsorption CD exploits the helical pitch of 
the laser field in \textbf{space}, PXCD  takes advantage of the 
sub-cycle rotation of the light field 
in \textbf{time} and is inherently ultrafast. Indeed, PXCD arises only if 
the excitation dipoles $\vec{d}_{01}$, $\vec{d}_{02}$ are non-collinear:
for the angle $\phi$ between the two transition dipoles, the PXCD (Eqs. (\ref{PXCD},\ref{PXCD_cur})) is proportional to 
$\sigma\sin(\phi)$. Since $\sigma=\pm 1$, $\sigma\sin(\phi)=\sin(\sigma\phi)=\sin(\sigma\omega\tau)$, where $\omega$ is light frequency and $\tau=\phi/\omega$ is the required time for the light field to rotate by the angle $\phi$. %Thus, PXCD relies on the memory effect: the helicity of the exciting 
%field is imprinted onto the coherent dynamics of the created chiral wave-packet.
PXCD vanishes if the coherence between excited states $|1\rangle$ and $|2\rangle$ is lost and reflects dynamical symmetry breaking in an isotropic medium. 

The oscillations of the PXCD signal 
Eqs.(\ref{PXCD},\ref{PXCD_cur}) appear to suggest that probing it requires the combination of ultrafast time resolution and chiral sensitivity. 
We now show that time-resolving PXCD does not, in fact, require a chiral probe. The coherence underlying PXCD allows a chiral object to 'interact with itself', albeit in a different quantum state, thus mimicking interaction with "another chiral object" and removing any need for other chiral interactions during the probe step.  One such non-chiral probe, termed PhotoeXcitation-induced photo-Electron Circular Dichroism (PXECD), is introduced below.

%We show that this assumption is not generally correct: PXCD can be probed with non-chiral light, linearly polarized  with polarization direction fixed in space, and with no help from any additional chiral interactions.
%On the other hand, we will now show that such probing requires 
%no help from other chiral interactions.
%The trick is that the coherence underlying PXCD allows us to make a chiral object to interact with itself.

%, such as circularly polarized pulses or "molecular helixes" responsible for
%creating the PXCD. 

 \section{Probing chiral dynamics in bound states}

 One way to probe the excited chiral density is to promote the chiral wave-packet to the electron continuum using a \textbf{linearly} polarized pulse (Fig 1c). As shown in the SI, the standard photoionization observable,  the  photoelectron current averaged over molecular orientations, is:
 %\beq
 %\label{PXECD}
 %j_z(|k|)=\sigma[\vec{d}_{01}\times \vec{d}_{02}]Re\vec{D}_{12}(|k|)\sin(\Delta E_{12}\tau)+\sigma[\vec{d}_{01}\times \vec{d}_{02}]Im\vec{D}_{12}(|k|)\cos(\Delta E_{12}\tau). 
 %\eeq
 \beq
 \label{PXECD}
 J_z^{PXECD}(k)=%\sigma[\vec{d}_{01}\times \vec{d}_{02}]\vec{D}_{12}^{*}(k)e^{i\Delta E_{12}\tau}+c.c.=\nonumber\\
 \sigma[\vec{d}_{01}\times \vec{d}_{02}]\vec{D}_{12}^{r}(k)\sin(\Delta E_{21}\tau)-\sigma[\vec{d}_{01}\times \vec{d}_{02}]\vec{D}_{12}^{i}(k)\cos(\Delta E_{21}\tau), 
 \eeq
with $J_x^{PXECD}(k)=J_y^{PXECD}(k)=0$.
Here   $\tau$ is the pump-probe delay, $\vec{D}_{12}(k)=\vec{D}_{12}^{r}(k)+i\vec{D}_{12}^{i}(k)$ is the Raman-type photoionization vector (see the SI) which connects the excited bound states via the common continuum and plays the role of $\vec d_{12}$ of Eq.(\ref{PXCD},\ref{PXCD_cur}) and $k$ is the photoelectron momentum.
%Other notations in Eq. (\ref{PXECD}) are the same as in Eq. (\ref{PXCD}).

First, the electron current Eq. (\ref{PXECD}) is proportional to the helicity $\sigma$ of the pump pulse. Second, as transitions to the continuum are described by complex dipole vectors, it contains two triple products. Just like the triple product 
$[\vec{d}_{01}\times \vec{d}_{02}]\vec{d}_{12}$ earlier, both $[\vec{d}_{01}\times \vec{d}_{02}]\vec{D}_{12}^{r}$ and $[\vec{d}_{01}\times \vec{d}_{02}]\vec{D}_{12}^{i}$  will change sign upon reflection. Thus, the electron current reverses its direction if the handedness of the pump pulse or of the enantiomer is swapped, showing that PXECD  is a genuine chiral effect.  %$j_z^{PXECD}(|k|)$ is proportional to the co-called forward-backward asymmetry in electron emission along the propagation direction and serves as a measure of chirality. 
%How does this asymmetry arise in photoionization by linear field? The helicity of light recorded in the dynamics of the excited  chiral wave-packet is  revealed when the wave-packet is probed by the linear field due to the interference of  transitions from the excited states to the same continuum state.  This interference reads out the information recorded during the excitation step (Fig.1 c). 
The chiral nature of the response arises only if the participating bound states are coherently excited.
Once the coherence is lost, the chiral signal will also disappear.

Importantly, the state of the continuum (Fig 1c) does not need to be chiral, as it only provides a link between the two chiral bound states.  $J_z^{PXECD}(k)$ remains chiral even for a plane wave continuum (see the SI), in this case $\vec{D}_{12}(k)$ only has an imaginary component:
 \beq
 \label{PXECD_PW}
 J_{z,PW}^{PXECD}(k)=%\sigma[\vec{d}_{01}\times \vec{d}_{02}]\vec{D}_{12}^{*}(k)e^{i\Delta E_{12}\tau}+c.c.=\nonumber\\
 -\sigma[\vec{d}_{01}\times \vec{d}_{02}]\vec{D}_{12}^{i,PW}(k)\cos(\Delta E_{21}\tau). 
 \eeq
%Moreover, the total photoelectron current is directly linked to the helical current excited in bound states $j_z^{PXCD}$ (Eq. \ref{PXCD_cur}) (see the SI).
The total photoelectron current $J_{tot}^{PXCD} = \int J_{z,PW}^{PXECD}(k)dk$ measures the helical current excited in bound states $j_z^{PXCD}$ (Eq. \ref{PXCD_cur}) distorted by the partial alignment of the molecular ensemble induced by the pump (see the SI).
%\beq
% \label{PXECD_PW}
% \int J_{z,PW}^{PXECD}(k)dk \propto %\sigma[\vec{d}_{01}\times \vec{d}_{02}]\vec{D}_{12}^{*}(k)e^{i\Delta E_{12}\tau}+c.c.=\nonumber\\
% \sigma[\vec{d}_{01}\times \vec{d}_{02}]\vec{\tilde d}_{12}\Delta E_{21}\cos(\Delta E_{21}\tau), 
% \eeq
%where $x,y$- Cartesian components of the dipole $\vec{\tilde d}_{12}$  are related to the ones of the molecular dipole $\vec{ d}_{12} $ as ${\tilde d}_{12}^{x,y}=\alpha {d}_{12}^{x,y}$, $0<\alpha<1$ (see the SI).   The probe pulse detects   "compression" of the dipole $\vec{ d}_{12} $ in the pump pulse polarization plane %is an instrumental response of our probe, 
%due to partial alignment of initially isotropic molecular ensemble upon excitation by the pump.
%\beq
% \label{PXECD_PW}
% \int J_{z,PW}^{PXECD}(k)dk \propto %\sigma[\vec{d}_{01}\times \vec{d}_{02}]\vec{D}_{12}^{*}(k)e^{i\Delta E_{12}\tau}+c.c.=\nonumber\\
% \sigma[\vec{\tilde d}_{01}\times \vec{\tilde d}_{02}]\vec{d}_{12}\Delta E_{21}\cos(\Delta E_{21}\tau), 
% \eeq
%where the dipoles $\vec{\tilde d}_{01}$, $\vec{\tilde d}_{02}$   are expressed via the molecular dipoles $\vec{d}_{01}$, $\vec{d}_{02}$  as ${\tilde d}_{01,02}^{j}=\delta_j {d}_{01,02}^{j}$, ($j=x,y,z$), $\delta_x=\delta_y=\delta_z^{-1}$.   The probe pulse sees the "stretching" of these dipoles in the pump pulse polarization plane ($\delta_{x,y}>1$) and "compression" in the propagation direction ($\delta_{z}=\delta_{x,y}^{-1}$) %is an instrumental response of our probe, 
%due to partial alignment of initially isotropic molecular ensemble upon excitation by the pump.
One might think that partial alignment of the excited molecular ensemble could already be fully responsible for enabling non-chiral probes of chiral dynamics. It is not true in our case. Indeed, the effect of alignment  persists for a single excited electronic state and for the two excited electronic states with collinear dipoles, but in both cases it leads to zero PXECD current. %Finally, once the effect of partial alignment is removed from Eq.(4), the total PXECD current is directly proportional to the chiral component of the helical current in bound states: $J_{tot}^{PXCD}\propto j_z^{PXCD}$ (see SI). 
Finally, removing the effect of partial alignment  from Eq.(4) shows that the  PXECD current remains chiral for every $k$, while $J_{tot}^{PXCD}$  becomes  directly proportional to the chiral component of the helical current in bound states: $J_{tot}^{PXCD}\propto j_z^{PXCD}$ (see the SI).

%In case of plane-wave continuum, the effect of alignment only slightly distorts the mapping between the bound and continuum currents (see SI).   

Probing the created chiral excitation using  photo-electron imaging with linearly polarized light  constitutes yet another new phenomenon, PhotoeXcitation-induced photoElectron Circular Dichroism (PXECD). PXECD is reminiscent of the Photoelectron Circular Dichroism (PECD) \cite{ritchie76,powis00, Bowering01,nahon15,garcia13,nahon16,comby2016relaxation}, which  arises when a circularly polarized light is used to photoionize a chiral molecule. However, there is a fundamental difference. %In PECD, the liberated electron interacts with the chiral molecular potential, which acts like a helix converting the induced current in the polarization plane into the motion along the light propagation direction and leading to enantio-sensitive sign of the current in this  direction. 
 PECD can only exist if the molecular potential felt by  the emitted electron is chiral \cite{ritchie76} (the effect becoming negligible for photoelectron energies above 10 eV), while the initial orbital may or may not be chiral at all \cite{ulrich2008giant}. It is also clear from the diagram of PECD in Fig 1 (d). The continuum state cannot merely serve  as a non-chiral link, as in this case it will only mediate the coupling of the chiral object, the molecule in the ground state, "to itself" rather than to another chiral object.% The need for significant interaction of the photoelectron with the chiral part of the molecular potential limits PECD to relatively slow photoelectrons ($\lesssim 10 eV$).

 %That is why PECD is usually detected relatively close to the ionization threshold (below 10 eV), where the photo-electron interaction with the chiral part of the molecular potential is still significant. 
 %In practice, disentangling the interplay of bound and continuum chirality in PECD is formidable task, as it requires accurate photoionization calculations for chiral molecules.
 
 In contrast to PECD, the PXECD  requires neither chiral light, nor chiral scattering of the photo-electron.  Since PXECD does not require the technically challenging use of 
 ultrashort circularly polarized XUV pulses \cite{wang12_femto,Spezzani11_coher,allaria12_highly,fleischer2014spin,ferrea.15}, it  opens unique perspectives for ultrafast chiral-sensitive measurements using readily available linearly polarized UV and XUV light from table-top high harmonic generation sources, with no restrictions on photoelectron energies.

We shall now confirm both numerically and experimentally that our scheme provides a sensitive time-resolved probe of chiral molecular dynamics in bound states.
%The ultrafast nature of interactions underlying PXCD and its  sensitive probe (PXECD) opens unique opportunities of ultrafast probing of molecular chirality. Controlling the time-delay between circular pump and linear probe pulses allows us to do just that: to time-resolve the characteristic asymmetry in the photoelectron distribution arising from coupled electronic and vibrational chiral dynamics. 

\section{Theoretical analysis in fenchone}

%In addition to the measured vibronic relaxation and the 1 psec internal dynamics, the wave-packet motion in fenchone should include even faster components associated with the quantum beating between \textit{s-} and the manifold of 3 \textit{p-} Rydberg states, with typical periods between $\sim$10 fs and $\sim$100 fs. These dynamics are too fast to be resolved with our current experimental temporal resolution ($\sim$170 fs), but can be investigated theoretically. 

To quantify the PXECD effect we performed  quantum mechanical calculations on fenchone molecules (see the SI). First, we simulated the PXCD phenomenon and calculated the excitation of the s- and p-manifold of Rydberg states in  fenchone by  a circular pump pulse. The resulting electron density of the Rydberg wave-packet is asymmetric in the $z$-direction in the momentum space. The asymmetry reverses if the helicity of the pump pulse or the handedness of the molecule is reversed.
 The strength of the PXCD can be quantified by the magnitude of the chiral component of the excited electron density. It is obtained by subtracting the momentum space density $D$  obtained with right (R) and left (L) polarized light: $PXCD=2(D(L)-D(R))/(D(L)+D(R))$.
Even after averaging over molecular orientations, the calculated PXCD reaches very high values (35$\%$, Fig. 2(a)). The asymmetry of the charge distribution corresponds to a macroscopic dipole moment $d_z^{\rm PXCD}$ which  reaches 3 Debye (Fig. 2(b)) and oscillates at frequencies determined by the energy differences between the states forming the electronic wave-packet. 
The  calculated pump-probe PXECD signal reveals these oscillations (Fig. 2c).  While few-femtosecond pulses would be needed to resolve them,
%providing a unique picture of the chiral electronic dynamics. However, 
the PXECD signal can also be detected with much longer pulses. Fig. 2(d) shows that both PXCD and PXECD  survive temporal averaging over 100 fs duration of a probe pulse. %, thanks to relatively slow dynamics in the p-manifold of the Rydberg states. 
%To demonstrate the  sensitivity of the 
%PXECD signal to pure bound state chirality,  
%we used a plane wave continuum  in our
%simulations.

\section{Observation of PXECD in fenchone}

%To draw out coherent excitation in excited states of chiral molecules we take advantage of Rydberg states, as their almost parallel potential electron surfaces should allow one to minimize possible decoherence due to nuclear motion.
In our experiment, a circularly polarized femtosecond pump pulse at 201 nm (6.17 eV photon energy, 80 meV at 1/e bandwidth) photoexcites enantiopure fenchone molecules from a supersonic gas jet in the interaction zone of a velocity map imaging spectrometer. The molecules are excited to their first (\textit{s-} and \textit{p-}) Rydberg bands through single-photon absorption (Fig. 3 (a), see the SI). A time-delayed, linearly polarized probe pulse at 405 nm (3.1 eV photon energy, 35 meV FWHM bandwidth) induces one-photon ionization of  the excited molecules. The cross-correlation of the pump and the probe pulses is 170 fs. The photoelectrons are  accelerated and projected by an electrostatic lens onto a set of dual microchannel plates and imaged by a phosphor screen and a CCD camera. 
The photoelectron images are recorded alternatively using left (LCP) and right (RCP) circularly polarized pump pulses. 
The difference (LCP-RCP) and sum (LCP+RCP) of these two images are reconstructed using a least-square fitting algorithm (see the SI).
 We define the PXECD signal as
$PXECD=\frac{2(LCP-RCP)}{(LCP+RCP)}$ and the photoelectron spectrum (PES) as $PES=(LCP+RCP)/2$. Both are shown in Fig. 3(b) for a 200 fs pump-probe delay. As expected, a significant PXECD signal is observed, reaching 1 $\%$ \cite{footnote3}
 in good agreement with our calculations (Fig. 2(d)).

%The Rydberg states of (1S)-(+)-fenchone populated in our experiment decay in 3.3 ps. 
%The PXECD results from the coherent superposition of several of such states, i.e. from a wave-packet with much faster dynamics due to quantum beating. 
%The ultrashort nature of the light pulses we used enables us to track these dynamics, with a 170 fs resolution determined by the cross-correlation of the pump and the probe pulses.
The photoelectron spectrum contains a single broad component, corresponding to ionization from the outermost orbital (vertical ionization potential ~8.72 eV). The position of this component does not shift with the pump-probe delay (Fig. 4 (b)) and decays in 3.3 ps, reflecting  simple vibronic relaxation of the Rydberg population onto lower states which cannot be photoionized by the 3.1 eV probe photons. The temporal evolution of the PXECD image shows much richer spectroscopic features, which can be analyzed by decomposing it in odd Legendre polynomials (Fig. 4(a)). We note that a sum of first- and third-order Legendre polynomials, with coefficients $\alpha$ and $\alpha'$, is enough to get the PXECD images. Both coefficients maximize around $\sim$ 50 meV below the maximum of the PES.
The  PXECD signal (Fig. 4(b)) can be decomposed into two components: below and above the maximum of the PES. The low-energy component of $\alpha$ undergoes a rather smooth decay. On the contrary, its high-energy component decays very quickly and even changes sign around 1 ps. For $\alpha'$  the
behaviour is opposite, \textit{i.e.} the high-energy component shows much slower dynamics than the low-energy part.
Such time- and electron energy- dependent behaviour is characteristic of internal vibrational torsional motion  and may indicate the change of the chiral structure of the molecule induced by such motion. Indeed, the electronic excitation of the molecules is expected to be accompanied by a significant vibrational excitation, since the equilibrium geometries of the 3s and 3p Rydberg states are quite different from that of the ground state. The molecules will tend to relax towards the equilibrium geometry of the Rydberg states, and oscillate around it. Figure 5 illustrates the influence of this change of molecular geometry on the calculated PXECD signal. %Even though the bond length changes are quite small ($\leq 7 \%$), a significant modification of the PXECD signal is observed.
Even small bond length changes ($\leq 7 \%$) lead to significant modification of the PXECD signal.
This demonstrates the remarkable sensitivity of PXECD to molecular vibrations, which  follow the electronic excitation.
At 4 ps (not shown), the PXECD completely vanishes while the Rydberg population is still significant. This result unambiguously reflects the loss of wave-packet coherence which halts chiral dynamics in our experiment.

\section{Vibrational PXCD: experiments in camphor}

Is it possible to create PXCD  from purely vibrational excitation of a chiral molecule?
Theoretically, the two excited states in Eqs.(1,2) needed for PXCD do not have to be different electronic states. Vibrational states within the same electronic state can also fulfil the PXCD condition as long as their dipoles are not collinear, see Eqs. (1,2).
As shown in the SI, this requires the breakdown of the Franck-Condon approximation, which is caused by a strong dependence of the electronic wave-function on the position of the nuclei. In turn, such dependence leads to the appearance of electronic currents stimulated by the nuclear motion, which is triggered by the pump pulse. Thus, vibrational PXCD is intertwined with the underlying chiral motion of electrons. Note that this strong dependence of the electronic wave-functions on the nuclear positions naturally arises in the vicinity of conical intersections between electronic potential surfaces. Thus, we expect that PXECD could be used to excite and reveal coherent chiral dynamics at conical intersections.

To gain further insight into the role of electronic versus vibrational dynamics in PXECD, we performed measurements in (1R)-(+)-camphor, a very similar structural isomer of fenchone. The \textit{s-} and \textit{p-} Rydberg bands of camphor are upshifted by additional several tens of meV compared to fenchone, preventing  direct  excitation of the \textit{p-} states and thus of an electronic chiral wave-packet. Nevertheless, the experiment still reveals a strong PXECD signal, indicating that a chiral vibronic wave-packet has been created in the \textit{s-} Rydberg band of camphor. The $\alpha'$ coefficients  in camphor and  fenchone are of opposite sign as seen in multiphoton \cite{Lux2015_photo} and one-photon PECD \cite{nahon15}.
%The $b_3$ coefficients  in camphor and  fenchone are of opposite sign \cite{Lux2015_photo,nahon15} as seen in multiphoton \cite{Lux2015_photo,lux12,lehmann13} and one-photon PECD \cite{nahon15}.
In our experiment, this could be a consequence of PXECD sensitivity to isomerism (see  Figure 5 to gauge the sensitivity to nuclear configuration), but it could also be a signature of the different nature of the excited chiral electronic currents in fenchone and camphor.
Changing the excitation wavelength from 202 nm to 200 nm does not affect the monoexponential decay of the PES. 
In contrast, a strong change is observed in the PXECD: the $\alpha'$ magnitude is almost twice as large and it is  shifted in energy towards the red wing of the photoelectron spectrum.
The drastic change observed in the PXECD signal in camphor once the pump photon  energy is increased by only 60 meV  illustrates the extreme sensitivity of this measurement to the excited vibrational dynamics.

\section{Conclusions and outlook}

We have demonstrated two new phenomena. 
First, we have shown the efficient excitation of a macroscopic bound chiral electron density in the excited states of randomly oriented chiral molecules without the help of magnetic interactions (the PXCD phenomenon).
%Thus, we have used non-chiral light to create a macroscopic chiral density in an isotropic ensemble of chiral molecules. 
In the dipole approximation the chiral pitch of circularly polarized light vanishes. This means that the creation of the macroscopic chiral density in the isotropic ensemble of chiral molecules is based not on the helical structure of light, but on its planar rotation.

Second, we have shown that the resulting chiral  dynamics  can be probed without the help of further chiral interactions and thus in an 
efficient and versatile way.
The  detection relies on photoelectron circular dichroism arising from the ionization of excited molecules by linearly polarized light pulses (the PXECD phenomenon), but is not limited to this scheme. The application of a linearly polarized XUV probe in PXECD would enable genuine probing of ultrafast chiral bound dynamics, since PXECD does not require chiral interaction in the continuum, which becomes negligible for sufficiently high-energy electrons. 

The ensemble-averaged chiral charge density arising in PXCD implies asymmetry in charge distribution along the light propagation direction. Depending on the medium density, this could lead to a very large coherently oscillating macroscopic dipole. The phase of this oscillation is opposite for two enantiomers, leading to macroscopic enantio-sensitive effects. The existence  of the  enantio-sensitive macroscopic dipole  opens the way to the separation of enantiomers in isotropic racemic mixtures in the gas phase.

%in the gas phase via application of additional electric fields \cite{filsinger2009}. 

The PXCD phenomenon opens the way to direct visualization of chiral electronic density using time-resolved X-ray diffraction imaging,  both in the gas and condensed phase. Intense ultrafast sources of X-ray radiation, such as Free Electron Lasers, combined with measurements, sensitive to valence-shell dynamics in the gas phase \cite{bredtmann2014x} should  lead to few-fs time resolution of chiral charge dynamics.% The  PXCD combined with such methods as femtosecond X-ray powder diffraction \cite{Elsaesser} could open a way to imaging chiral charge dynamics in the condensed phase.   

Finally, PXCD could be used to drive  molecular reactions in chiral systems in a stereo-specific way, by imprinting a chiral torque via the helicity of the exciting circularly polarized pulse. The ultrafast charge dynamics triggered by coherent electronic excitation is reminiscent of ultrafast charge migration triggered by photo-ionization \cite{Lunnemann,Breidbach,Remacle,Kuleff,Lepine,Leone,Kuleff2} recently observed in ref. \cite{Calegari}  and  speculated  to underlie 
charge-directed reactivity in cations \cite{Weinkauf}. 
Chiral electron stereo-dynamics  in neutral molecules  may open similar opportunities for controlling charge and energy flow in molecules at the level of electrons, offering new perspectives for such intriguing problems as asymmetric synthesis, a major challenge in stereochemistry.  

\begin{methods}
An Even-Lavie valve is used as a pulsed enantiopure fenchone source with helium as carrier gas to avoid cluster formation. (1R)-(–-) and (1S)-(+)-fenchone correspond to (1R,4S) and (1S,4R) fenchone respectively.  The 170 fs cross-correlation time as well as the 0 fs delay are determined on the lightest fragment C$_4$H$_5$O$^+$ produced by dissociative ionization with both linearly polarized pump and probe. The high voltage of the repeller electrode was -3kV for the experiment done in fenchone and only -2kV for the experiment done in Camphor, which increases the energy resolution. Note that the latter, along with the energy calibration, that has been determined by photoionizing krypton. Typically the energy resolution is 80 meV at 0.7 eV kinetic energy. The presented results are obtained by scanning the pump-probe delays typically 30 times. At each delay, helicity is permuted each 45000 laser shots (=45 seconds)  to record several images. %The normalized momentum distribution of the photoelectrons (ARPES(L)+ARPES(R)) and (ARPES(L)-ARPES(R)) for one enantiomer is recovered as a sum of Legendre polynomials by a pBasex analysis of the images. The even b$_i$ coefficients encode the symmetric image (ARPES(L)+ARPES(R)), while the odd ones, the asymmetric image (ARPES(L)-ARPES(R)).
\end{methods}
%begin{thebibliography}{1}
%end{thebibliography}
%\bibliography{biblio_august}

\begin{thebibliography}{10}
\bibitem{footnote} For non-magnetic materials, light absorption  is proportional to the scalar product of the  induced polarization current and electric component of the light field. Thus, even if the motion of bound electrons underlying the induced current were helical, moving out of the polarization plane of the circular light, it would have not contributed to the CD signal. For magnetic materials light absorption includes an additional term: a scalar product between magnetic field vector and the time-derivative of induced magnetization. It does not change our conclusion: CD is only sensitive to electron dynamics in light polarization plane also for magnetic materials.
\bibitem{footnote3}The Forward-Backward Asymmetry, i.e. the normalized difference of photoelectron counts in forward and backward hemispheres relative to z axis, is 0.62$\%\pm$ 0.03$\%$.
\end{thebibliography}
%\end{document}

\section{Acknowledgements}
We thank Rodrigue Bouillaud and Laurent Merzeau for technical assistance.
We thank M. Ivanov, A. Stolow and T. Elsaesser for stimulating discussions.
We acknowledge financial support of the Agence Nationale pour la Recherche (ANR-14-CE32-0014 MISFITS) and the University of Bordeaux. Z.M. and O.S. gratefully acknowledge the support from Deutsche Forschungsgemeinschaft, project Sm 292-5/1, A.H. gratefully acknowledges the support from Deutsche Forschungsgemeinschaft, project Iv 152/7-1. A.F.O. and O.S. gratefully acknowledge EU ITN MEDEA -AMD-641789-17 project. S.B. acknowledges the support of a NSERC Vanier Canada Graduate Scholarship. R.G. acknowledges financial support from the Agence Nationale pour la Recherche through the XSTASE project (ANR-14-CE32-0010).

\section{Additional information}
\textbf{Competing Interests} The authors declare that they have no competing financial interests. Requests for materials and correspondence should be addressed to valerie.blanchet@celia.u-bordeaux.fr, yann.mairesse@celia.u-bordeaux.fr, Olga.Smirnova@mbi-berlin.de, bernard.pons@u-bordeaux.fr

\section{Author contributions}
S.B., A.C., R.G., Y.M., V.B., performed the experiment. D.D. and S.P. operated the laser system. S.B., A.C., B.F., G.G., L.N., B.P., Y.M. and V.B. analyzed the data. B.F. and B.P. performed the molecular geometry and dynamical calculations. A.H., Z.M. and O.S. developed the analytical theory, A.F.O. and O.S. derived  chirality measures for PXCD and PXECD. S.B. wrote the first version of the manuscript, all authors contributed to writing the manuscript.

\newpage
\textbf{FIGURE 1}: Chiral discrimination schemes. C.C. denotes complex conjugated, i.e. time-reversed process. Downarrows denote C.C. of driving fields.
(a) CD requires magnetic dipole transition up and electric dipole transition down and vise-versa. (b) PXCD (Eq.\ref{PXCD}) requires coherent excitation of two states by ultrashort circularly polarized pulse. The stimulated dipole transition to state $|2>$ is followed by  dipole transition to state $|1>$ and stimulated dipole transition to state $|0>$. Insert: Induced dipole maps out a helix as a function of time. (c) In PXECD (Eq.\ref{PXECD}) the two excited states are connected by Raman-type transitions via continuum, stimulated by linearly polarized pulse. (d) PECD requires circularly polarized light and photoelectron scattering off chiral potential $V_{ch}$. \\

\begin{figure}
\begin{center}
\includegraphics[width=12 cm,keepaspectratio=true]{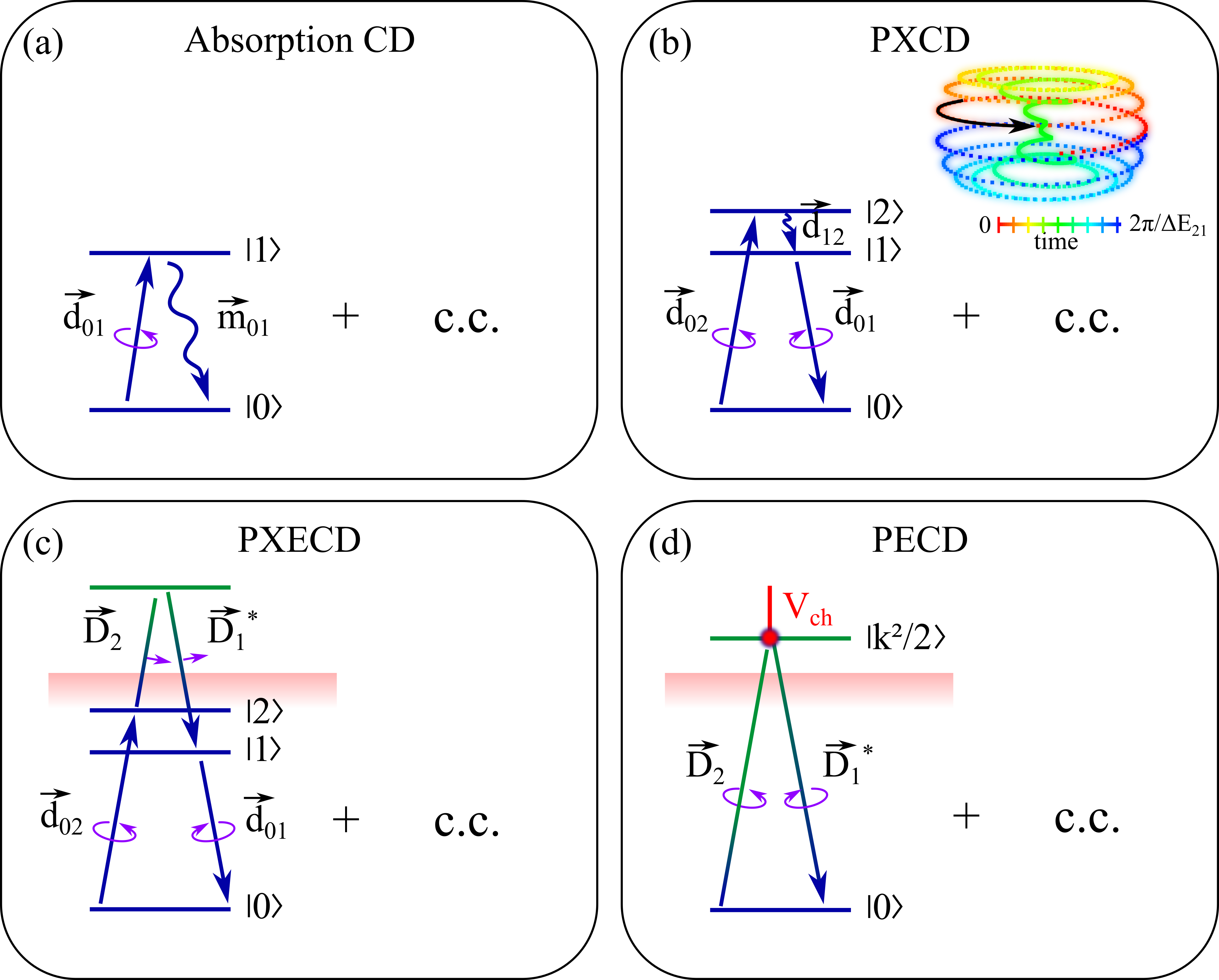}
\label{fig1}
\end{center}
\end{figure}

\newpage 
\textbf{FIGURE 2}: Theoretical analysis of  electron chiral dynamics in (1S)-(+)-fenchone.
(a) Momentum space electron density underlying PXCD. The asymmetry signifying chirality of the electron density is formed in light propagation direction z. (b) Temporal evolution of the x-, y- and z-components of the macroscopic dipole associated with the (3s,3p) Rydberg wave-packet created by a pump pulse (shaded area). Only the z-component, along the direction of propagation of the pump and probe pulses, survives orientational averaging. (c) Momentum space PXECD signals at various pump-probe delays $t$.
(d) Time-averaged PXECD to account for the temporal resolution of the experiment.\\
\begin{figure}
\begin{center}
\includegraphics[width=15 cm,keepaspectratio=true]{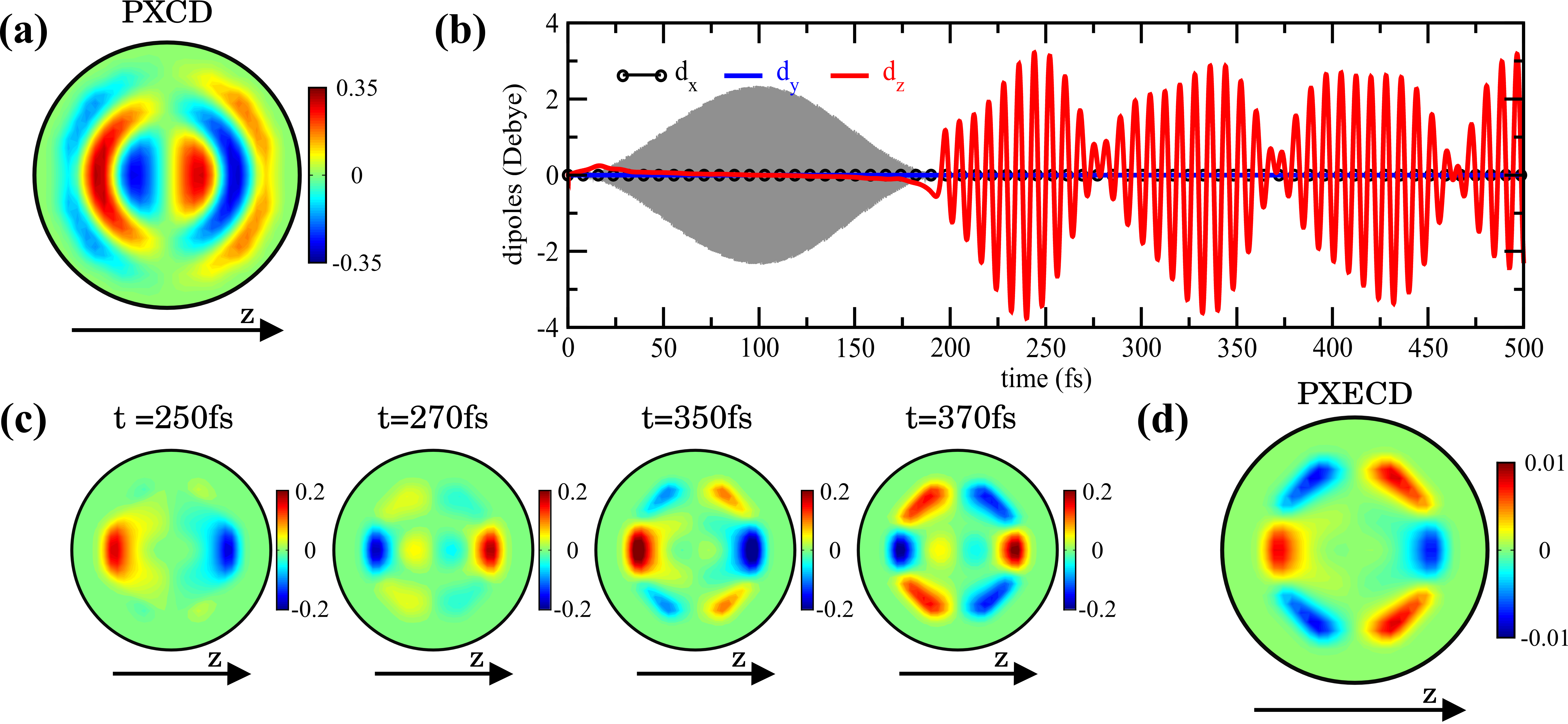}
\label{fig2}
\end{center}
\end{figure}

\newpage
\textbf{FIGURE 3 :} PXECD in fenchone molecules. (a) Absorption of circularly polarized pulse at 201 nm with helicity $\sigma=\pm1$ promotes an electron from the highest occupied molecular orbital to the \textit{s-} and \textit{p-} Rydberg bands, creating a chiral electron wave-packet. A linear probe pulse at 405 nm photoionizes the molecule, revealing the chiral asymmetry of the Rydberg wave-packet in the angular distribution of the photoelectrons. The absorption spectrum of fenchone is adapted from \cite{pulm97} (b) Experimental image of photoelectron spectrum (PES) and  PXECD images at 200 fs pump-probe delay for  (1S)-(+)-fenchone. The characteristic forward-backward asymmetry is observed in light propagation direction z.\\
\begin{figure}
\begin{center}
\includegraphics[width=8.7 cm,keepaspectratio=true]{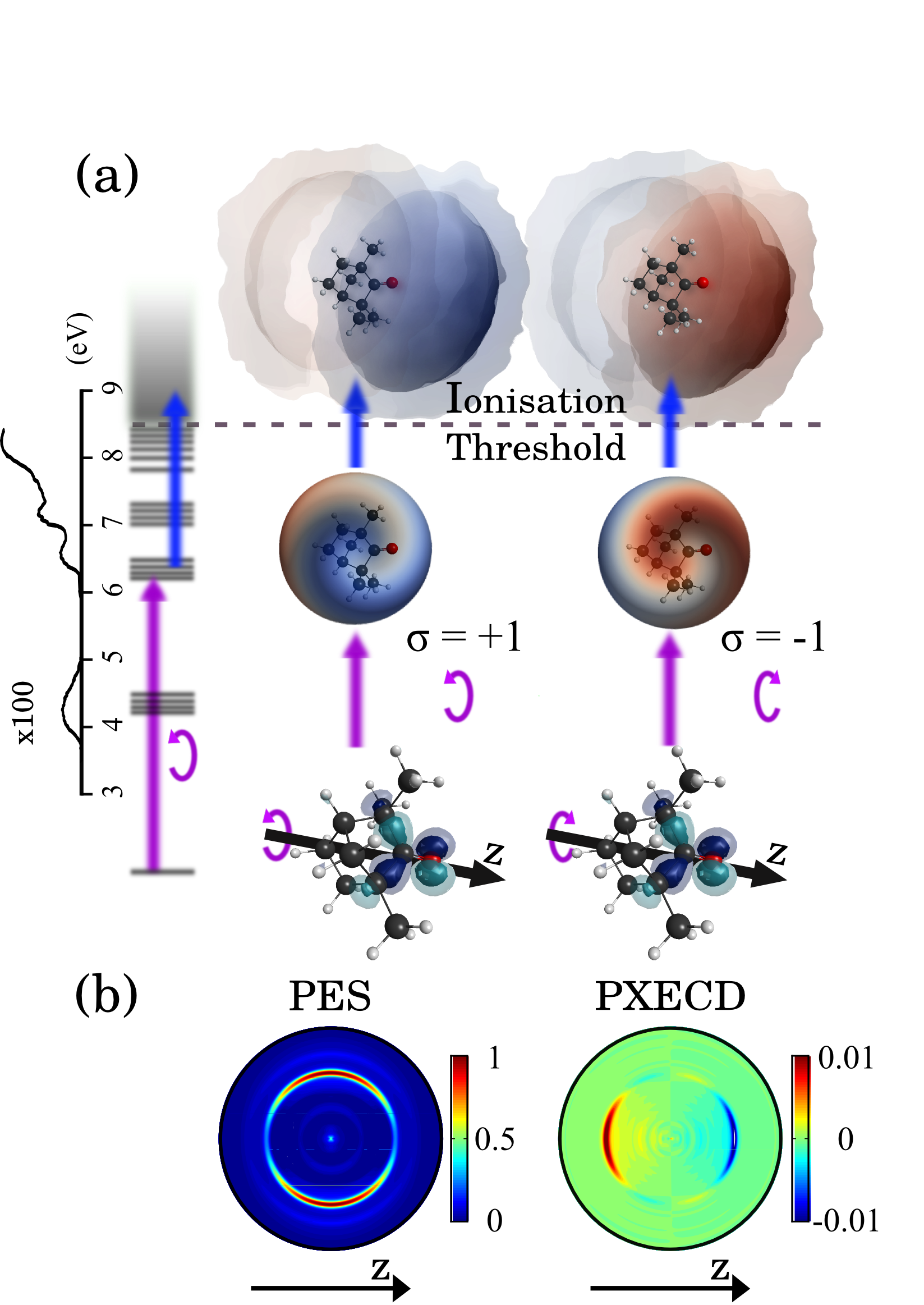}
\label{fig3}
\end{center}
\end{figure}

\newpage 
\textbf{FIGURE 4}: Time-Resolved PXECD in fenchone and camphor. (a) Legendre polynomials decomposition of the PXECD image for (1S)-(+)-Fenchone at 200 fs pump-probe delay. The $\alpha,\alpha'$ coefficients are multiplied by their associated Legendre polynomials $P_i(\theta)$: $P_1=cos(\theta)$, $P_3=(5/2\cdot cos^3(\theta)+3/2\cdot cos(\theta))$. (b) Evolution of the PES  and PXECD  coefficients  as a function of pump-probe delay, in (1S)-(+)-fenchone with 201 nm pump, (1R)-(+)-camphor with 202 nm pump and with 200 nm pump. The black dotted lines represent energies corresponding to the maximum of PES. $\alpha$ and $\alpha'$ are normalized to the  maximum value of PES.\\
\begin{figure}
\begin{center}
\includegraphics[width=11 cm,keepaspectratio=true]{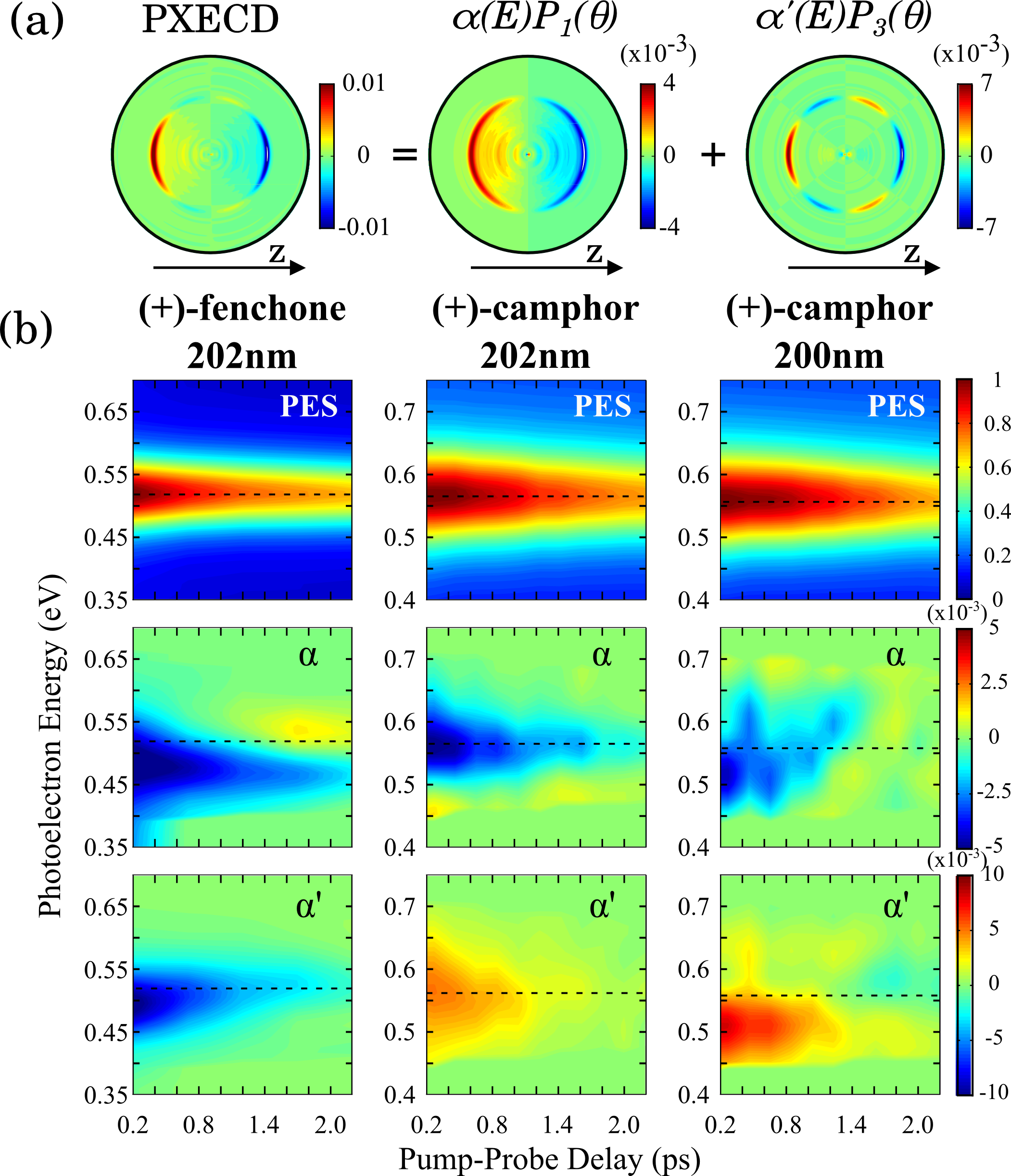}
\label{fig5}
\end{center}
\end{figure}

\newpage 
\textbf{FIGURE 5}: Sensitivity of PXECD in (1S)-(+)-fenchone to the evolution of the chiral molecular structure. (a)  The equilibrium geometries of the ground (dark) and 3s (shadowed) electronic states, and
the PXECD signal computed at the ground state geometry for both pump excitation and probe ionization. (b) The representation of the geometries are exchanged and the PXECD is computed assuming that pump excitation occurs at the ground state geometry while probe ionization occurs once the vibronic wave-packet has reached the 3s equilibrium geometry. The  PXECD image is averaged over random molecular orientations and the 100 fs duration of the probe pulse.\\
\begin{figure}
\begin{center}
\includegraphics[width=10 cm,keepaspectratio=true]{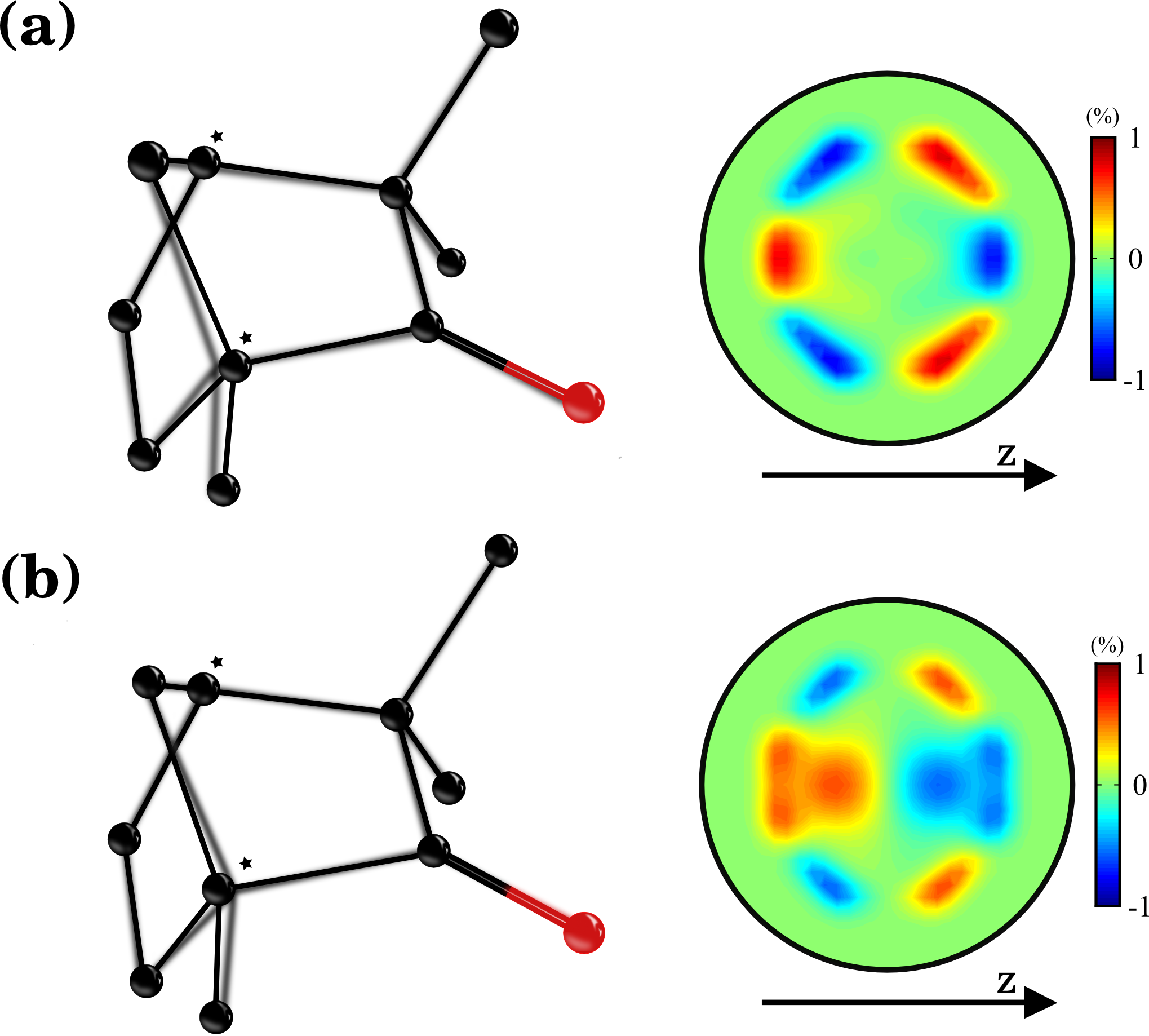}
\label{fig4}
\end{center}
\end{figure}
\end{document}